 \documentclass[
  aps,
  pre,
  onecolumn,
  superscriptaddress,
  citeautoscript
]{revtex4-2}
\setcitestyle{super}
\usepackage{lipsum}
\usepackage{mathrsfs}
\usepackage{bm,amsbsy,amssymb,amsmath}
\usepackage{caption}
\usepackage{subcaption}
\usepackage{graphics,graphicx,dcolumn,fleqn,epic,eepic,float,tabularx}
\usepackage{multirow,rotate,rotating,color}
\usepackage[utf8]{inputenc}
\newcommand{\figref}[1]{Fig.~\ref{fig:#1}}
\newcommand{\eqnref}[1]{Eq.~(\ref{eq:#1})} 
  \definecolor{tuered}{RGB}{214,0,74}
  \definecolor{tueblue}{RGB}{0,102,204}
  
  \newcommand{\revisedtext}[1]{\textcolor{black}{#1}}

\usepackage{tikz,pgfplots}
\usepackage{relsize}
\tikzset{fontscale/.style = {font=\relsize{#1}}}
\usetikzlibrary{calc}
\graphicspath{{img/}}

\usepackage{soul}
\usepackage{hyperref}
\begin{document}
\title{Free-Energy Analysis of Bubble Nucleation \\ on Electrocatalytic Surfaces}
  \author{Qingguang Xie}
   \email{q.xie@fz-juelich.de}
\affiliation{Helmholtz Institute Erlangen-N\"urnberg for Renewable Energy (IET-2),
Forschungszentrum J\"ulich, Cauerstra{\ss}e 1, 91058 Erlangen, Germany}

  \author{Paolo Malgaretti}
\affiliation{Helmholtz Institute Erlangen-N\"urnberg for Renewable Energy (IET-2),
Forschungszentrum J\"ulich, Cauerstra{\ss}e 1, 91058 Erlangen, Germany}

 \author{Othmane Aouane}
\affiliation{Helmholtz Institute Erlangen-N\"urnberg for Renewable Energy (IET-2),
Forschungszentrum J\"ulich, Cauerstra{\ss}e 1, 91058 Erlangen, Germany}

\author{Simon Thiele}
   \affiliation{Helmholtz Institute Erlangen-N\"urnberg for Renewable Energy (IET-2),
Forschungszentrum J\"ulich, Cauerstra{\ss}e 1, 91058 Erlangen, Germany}
\affiliation{Department of Chemical and Biological Engineering, Friedrich-Alexander-Universit\"at Erlangen-N\"urnberg, Egerlandstr. 3, 91058 Erlangen, Germany}

\author{Jens Harting}
   \email{j.harting@fz-juelich.de}
\affiliation{Helmholtz Institute Erlangen-N\"urnberg for Renewable Energy (IET-2),
Forschungszentrum J\"ulich, Cauerstra{\ss}e 1, 91058 Erlangen, Germany}
\affiliation{Department of Chemical and Biological Engineering and Department of Physics, Friedrich-Alexander-Universit\"at Erlangen-N\"urnberg, Cauerstra{\ss}e 1, 91058 Erlangen, Germany}

\begin{abstract}
 Bubble nucleation at catalyst surfaces plays a critical role in the operation of electrolyzers. However, achieving controlled bubble nucleation remains challenging due to limited understanding of the underlying mechanisms. Here, we present a free-energy model that quantitatively predicts both the activation energy and critical nucleus size of bubbles at given supersaturation, temperature, pressure, and surface wettability. 
We find that the activation energy $\Delta G_{max}$ decreases with increasing supersaturation $\zeta$, following a power-law scaling of $\Delta G_{max} \sim \zeta^{-2}$, while the critical nucleus radius $R_c$ scales as $R_c\sim \zeta^{-1}$. 
Our theoretical predictions for the critical nucleus radius of hydrogen, oxygen and nitrogen bubbles are in quantitative agreement with experimental measurements. 
Finally, we present a simple model that couples gas diffusion and electrochemical reaction kinetics to determine the maximum gas supersaturation at a given current density. Our results advance the fundamental understanding of bubble nucleation at catalyst surfaces and provide practical guidelines for catalyst layer design to improve the performance of electrolyzers.
\end{abstract}

\maketitle

\section{Introduction}
The nucleation of bubbles at catalyst surfaces is a common phenomenon in electrochemical devices for hydrogen production, such as water electrolyzers~\cite{yuan2023,deng2025,vincent2018,Ivanova2023TechnologicalPathwaysProduce}. 
The overall efficiency of electrolyzers is strongly influenced by bubble nucleation and the subsequent processes of bubble growth, departure, and transport. In proton exchange membrane water electrolyzers (PEMWE), for instance, oxygen bubbles may nucleate at the nanoporous anode catalyst layer (CL), migrate through the porous transport layer (PTL), and are eventually removed by convective water flow in the flow channels~\cite{yuan2023,YUAN2024,Mo2016,Jinke2017}. 
However, \revisedtext{the} bubbles growing at the catalyst layer may block active sites for oxygen evolution, leading to an increase in the overpotential and thus reducing the electrochemical performance of the device.
Understanding and controlling bubble nucleation at nano- and microscales is therefore essential for optimizing CL design and improving the overall electrolyzer efficiency.

Significant efforts have been made to understand the formation of bubbles at the catalyst layer. However, direct observation of nanobubbles is challenging due to the nanoscale, complex, and opaque nature of the porous material. \revisedtext{Furthermore, molecular dynamics simulations of bubble nucleation~\cite{Lohse2016, Perez2019, Lohse2024, ZHAN2025160} can provide nanoscale insights. However, they are typically restricted to specific parameters and small system sizes due to high computational cost. As a result, it is challenging to explore the full parameter space relevant to experiments or to capture bubble behavior in larger, more complex systems.
} These limitations have even led to ongoing debate about whether bubbles nucleate inside the CL or at the CL–PTL interface~\cite{Mo2016,Jinke2017,YUAN2024}. For example, Mo et al.~\cite{Mo2016,Jinke2017} observed bubble formation only on the CL surface adjacent to the PTL in PEMWE, challenging the assumption of uniform reaction activity. They attributed this to the high in-plane electrical resistance of the CL, which confines oxygen evolution to the CL–PTL interface. In contrast, Yuan et al.~\cite{YUAN2024} reported evidence of nanobubble accumulation and transport within the CL. They observed sustained bubble growth on the CL surface even in the absence of oxygen evolution, as well as the formation of gas outlet pores at the surface.
 
Theoretical predictions of bubble nucleation characteristics, including the activation energy and critical nucleus size of bubbles, may help to resolve these discrepancies. 
Since the early 20th century, numerous studies have sought to theoretically analyze the activation energy and critical nucleus size of bubbles both in the bulk liquid and at solid surfaces using classical nucleation theory~\cite{VolmerWeber1926,Farkas1927,KELTON201085,Lubetkin1995,Vachaparambil_2018}, yet these predictions have largely remained qualitative~\cite{Meloni2016}. \revisedtext{Alternatively, classical density functional theory (DFT)~\cite{Oxtoby1988,Shen2001} has been employed to study bubble nucleation by resolving the spatial variation of liquid density and microscopic molecular interactions. Nevertheless, it is computationally demanding, typically limited to small system sizes, and sensitive to the choice of approximate functionals.} Recently, Zargarzadeh et al. conducted a thermodynamic analysis of nucleating surface nanobubbles, considering closed systems with a finite number of molecules~\cite{zargarzadeh2016}. This work was later extended to a more quantitative framework by Lan et al.~\cite{lan2025}. Very recently, Gadea et al.~\cite{gadea2023} extended these efforts to explore nanobubble formation in open environments. However, these analyses rely on the number of molecules as an input parameter, which is difficult to measure experimentally. Additionally, a direct link between key parameters (e.g., supersaturation) and electrochemical reactions is still lacking in these models.

In this work, we consider bubble nucleation in an unbounded volume of liquid-gas solution, reflecting the work environment of electrolyzers. The system is maintained at constant temperature and pressure, with negligible total volume change. 
We perform a free energy analysis of bubble nucleation to quantitatively predict the activation energy required to nucleate a bubble with a critical radius at a catalyst surface under given conditions (e.g., pressure, temperature, supersaturation). We show that the wettability of the reactive surface largely affects the activation energy, whereas the critical nucleation size of bubbles remains constant, consistent with previous research~\cite{KELTON201085,Lubetkin1995,Vachaparambil_2018}. 
We find that the activation energy $\Delta G_{max}$ decreases with increasing supersaturation $\zeta$, following a power-law scaling of $\Delta G_{max} \sim \zeta^{-2}$, while the critical nucleus radius $R_c$ scales as $R_c\sim \zeta^{-1}$. 
Furthermore, we compare theoretical predictions of the critical nucleation radius for hydrogen, nitrogen and oxygen bubbles with experimental results~\cite{Edwards2019, Soto2018} and find quantitative agreement.  We then present a simple theoretical model to predict the maximal gas supersaturation achievable at a given current density. In addition, we propose possible descriptions for gas production, bubble formation, and transport within the catalyst layer and at the CL-PTL interface, which may explain the experimentally controversial reports~\cite{Mo2016,Jinke2017,YUAN2024}. \revisedtext{Overall, our models advance prior approaches by linking nanoscale bubble nucleation directly to experimentally relevant parameters. Unlike previous analyses that rely on the number of molecules and lack a connection to electrochemical reactions, we present a free-energy model that predicts activation energy and critical nucleus size as functions of supersaturation, temperature, pressure, and surface wettability. Additionally, a simplified kinetic framework couples gas diffusion with electrochemical reaction rates, enabling estimation of maximum supersaturation and nanoscale nucleation characteristics from operating conditions. By bridging nucleation thermodynamics and electrochemical kinetics across nano- to microscale, this framework provides predictive insight into bubble evolution relevant to electrocatalysis, energy conversion, and nanofluidic systems.
}

\section{Free energy analysis}
\label{sec:analysis}
\begin{figure}[h]
    \centering
        \begin{subfigure}{.32\textwidth}
		\includegraphics[width= 0.95\textwidth]{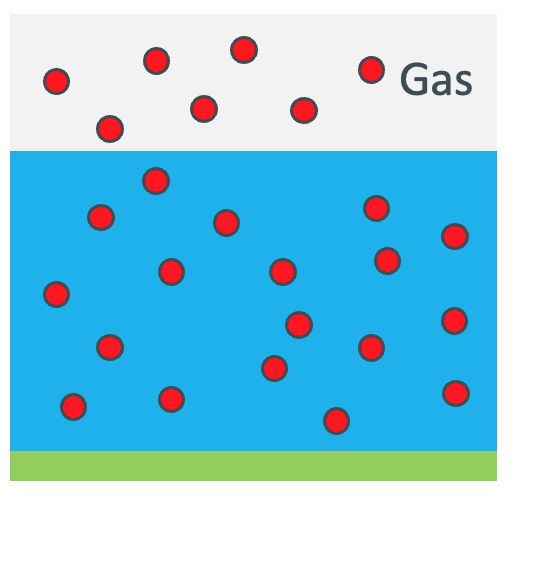}
		\subcaption{No bubble}
		\label{fig:snap-no-bubble}
	\end{subfigure}
 \hspace*{5mm}
	\begin{subfigure}{.363\textwidth}
		\includegraphics[width= 0.95\textwidth]{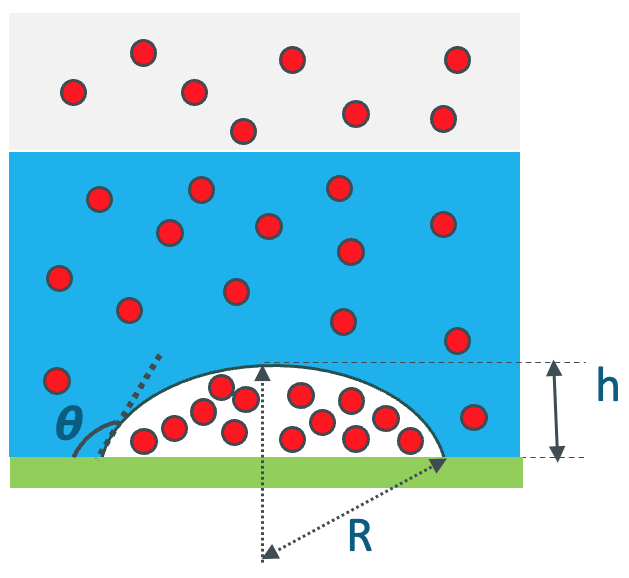}
		\subcaption{Single bubble}
		\label{fig:snap-single-bubble}
	\end{subfigure}
    \caption{Sketch of a system with a mixture of gas molecules and water without a bubble (a) and with a bubble (b). The red dots represent the gas molecules, the blue area represents water, and the white 
    area denotes the bubble. The solution is in contact with a gas phase, depicted in grey. The contact angle, radius, and height of the bubble are denoted as $\theta$, $R$, and $h$, respectively.
    }
    \label{fig:snap-bubble}
\end{figure}

We perform a free energy analysis of bubble nucleation at a catalyst surface to predict the energy difference between the states before and after nucleation. The system consists of a solution of water--or any solvent--and oxygen--or any gas--in contact with a solid surface at pressure $P$, temperature $T$, as illustrated in~\figref{snap-no-bubble}. Furthermore, the system is in contact with a gas phase, where the gases are in equilibrium. We assume that the number of gas molecules in the bubble is much smaller than the total number of gas molecules in the water, since the bubble volume is much smaller than the total water volume. In addition, the solution is assumed to be isothermal, and its volume remains constant during bubble formation. 

In the following, we assume that during the nucleation process, the total number of oxygen molecules in the system (i.e., solution plus bubble), $N$, is conserved. Under these circumstances, the relevant thermodynamic potential is the Gibbs free energy, written as  
\begin{equation}
    G_1(N,P) = N \mu_l(P,n_1) + \gamma_{ws} S_{ws},
\end{equation}
where $n_1$ is the number density of the oxygen molecules in the water before bubble nucleation,  $\mu_l$ is their chemical potential at given pressure and number density, $\gamma_{ws}$ is the surface energy of the water-solid interface, and $S_{ws}$ is the area of the water-solid interface. 

After bubble formation, we assume that $N_g$ molecules form a bubble of radius $R$\revisedtext{. The bubble} adopts a spherical cap shape on the solid surface in the limit of a low Bond number $Bo=\frac{\Delta \rho g R^2}{\gamma_{wg}}\ll 1$, where $\gamma_{wg}$ is the water-gas surface energy as shown in~\figref{snap-single-bubble}.
Here, $\Delta \rho$ is the density difference between water and the oxygen bubble,
and $g$ is the gravitational acceleration. 
The free energy after bubble formation is
\begin{eqnarray}
        G_2(N,P) &=& 2 \pi \gamma_{wg} R^2 (1+\cos \theta) + (N-N_g) \mu_l(P,n_2) + N_g\mu_g(P_b) 
        + \gamma_{ws} (S_{ws}-S_{gs}) + \gamma_{gs} S_{gs} ,
        \label{eq:G2}
\end{eqnarray}
where $\theta$ is the contact angle of the bubble on the catalyst surface, $n_2$ is the number density of the oxygen molecules in the water after bubble nucleation, $\mu_g(P_b)$ is the chemical potential of oxygen molecules in the bubble, and $S_{gs}$ is the footprint area of the bubble on the solid surface. $\gamma_{ws}$ and $\gamma_{gs}$ are the surface energies of the water–solid and gas–solid interfaces, respectively.
The free energy difference $\Delta G = G_2 -G_1$ reads
\begin{eqnarray}
    \Delta G  =  2 \pi \gamma_{wg} R^2 (1+\cos \theta ) + N \Delta\mu_l +N_g \Delta \mu_g + (\gamma_{gs}-\gamma_{ws}) S_{gs},
    \label{eq:DG}
\end{eqnarray}
where $\Delta \mu_l = \mu_l(P,n_2) - \mu_l(P,n_1)$ is the change in the chemical potential of the dissolved gas molecules, and $\Delta \mu_g = \mu_g(P_b)-\mu_l (P,n_2)$ is the chemical potential difference between the gas molecules in the bubble and those in the surrounding solution.
With $\cos\theta = \frac{\gamma_{gs}-\gamma_{ws}}{\gamma_{wg}}$ based on the Young's equation, and $S_{gs}=\pi (R\sin\theta)^2$, we get 
\begin{eqnarray}
        \Delta G &=&  2 \pi \gamma_{wg} R^2 (1+\cos \theta) + \pi \gamma_{wg}  R^2 \cos \theta  \sin^2 \theta + N \Delta\mu_l  +N_g \Delta \mu_g 
        \label{eq:energy_diff}
\end{eqnarray}
In the following, we assume that the oxygen solution is diluted such that the chemical potential of oxygen molecules in gas and in solution can be written as~\cite{solymosi2022} 
\begin{eqnarray}
    \mu_g(P)&=&k_B T \ln\left(\frac{P}{k_B T}\Lambda^3\right) \\
    \mu_l(P, n)&=& \mu^0_l(T)+k_B T \ln\left(n\Lambda^3\right).
\end{eqnarray}
Here, we assume the chemical potential of the oxygen in the gas bubble to be like an ideal gas. $\Lambda$ is the thermal wavelength, and $\mu^0_l(T)$ is the
chemical potential of oxygen in a specific, defined standard state, serving as a reference point. 

When a bubble forms, the overall volume of the system changes due to two contributions: an increase in volume from the formation of the bubble and a reduction in the volume of the liquid solution as fewer oxygen molecules remain dissolved.
In the limit of $N_g \ll N$ and neglecting the change in volume of the liquid solution, $\frac{N_G}{N}\gg \frac{\Delta v}{V}$, we obtain 
\begin{eqnarray}
    \Delta \mu_l &=& k_B T \ln\left(n_2\Lambda^3\right) - k_B T \ln\left(n_1\Lambda^3\right) \approx - k_B T \frac{N_g}{N}  \label{eq:deltal} \\
     \Delta \mu_g &=& k_B T \ln\left(\frac{P_b}{k_B T}\Lambda^3\right)-k_B T \ln\left(n_2\Lambda^3\right)-\mu_l^0 = k_B T \ln\left(\frac{P_b}{k_B Tn_2}\right)-\mu_l^0.
     \label{eq:deltag}
\end{eqnarray}
The number of molecules $N_g$ in the gas bubble can be determined by the equation of state of an ideal gas as
\begin{eqnarray}
        N_g &=& \frac{V}{k_B T} \left( P+\frac{2\gamma_{wg}}{R} \right) \nonumber \\ 
    &=& \frac{\pi R^3}{3k_B T} \left( P+\frac{2\gamma_{wg}}{R}\right) (2-\cos \theta)(1+\cos\theta)^2 ,
    \label{eq:ng}
\end{eqnarray}
where $P_b = P + \frac{2\gamma_{wg}}{R}$ according to the Young–Laplace equation,
and the volume of the spherical-cap-shaped bubble is taken as $V = \frac{\pi R^3}{3}(2-\cos \theta)(1+\cos\theta)^2$. 
Inserting Eqns.~\ref{eq:deltal}-\ref{eq:ng} into~\eqnref{energy_diff}, we obtain
\begin{eqnarray}
        \Delta G &\approx& 2 \pi \gamma R^2 (1+\cos \theta) + \pi \gamma  R^2 \cos \theta  \sin^2 \theta \nonumber \\
        &&- \frac{\pi R^3}{3} \left( P+\frac{2\gamma}{R} \right)(2-\cos \theta)(1+\cos\theta)^2
   \left(1-\ln\left(\frac{P+\frac{2\gamma}{R}}{k_B T n_2}\right)-\frac{\mu_l^0}{k_BT} \right) .
        \label{eq:energy_diff2}
\end{eqnarray}
Here, $\gamma=\gamma_{wg}$. By combining the $R^2$ and $R^3$ terms separately, and unfolding the $\theta$ terms, we get
\begin{equation}
    \Delta G \approx \frac{\pi}{3} (2+3\cos \theta -\cos^3 \theta) \left( \gamma R^2   \left[ 1 + 2\ln\left(\frac{P+\frac{2\gamma}{R}}{k_B T n_2}\right)+2\frac{\mu_l^0}{k_BT} \right] - R^3P\left[ 1- \ln\left(\frac{P+\frac{2\gamma}{R}}{k_B T n_2}\right)-\frac{\mu_l^0}{k_BT} \right] \right).
    \label{eq:energy_diff3}
\end{equation}
The expression above is a function of the number density, $n_2$, of dissolved oxygen \textit{after} bubble nucleation, whereas we would like to express it as a function of the number density (or preferably, the supersaturation) of the oxygen \textit{before} bubble nucleation, $n_1$. Accordingly, due to mass conservation, we recall that 
\begin{equation}
    n_2 V_2 + N_g = N .
\end{equation}
Due to our assumptions, $N_g\ll N$ and $V_2\simeq V$, we have $ n_2 \simeq n_1$.
Next we aim to relate $n_1$ to the supersaturation $\zeta$, i.e. the ratio between $n_1$ and the number density dissolved oxygen at equilibrium, $n_1^{eq}$. 
At equilibrium, the chemical potential of the oxygen in the gas phase equals that of the oxygen diluted in the solution, which leads to 
\begin{equation}
    k_BT\ln\left(\frac{P_{a}}{k_BT}\Lambda^3\right)=\mu_l^0 + k_BT\ln\left(n_1^{eq}\Lambda^3\right) ,
\end{equation}
where $P_{a}$ is the partial pressure of oxygen in the gas phase. Accordingly we get 
\begin{equation}
    n_1^{eq} = \frac{P_{a}}{k_BT}e^{-\frac{\mu_l^0}{k_BT}}\,.
    \label{eq:n_eq}
\end{equation}
Finally, using $n_2 \simeq n_1 = \zeta n_1^{eq} $, we obtain
\begin{align}
    \Delta G &\approx \frac{\pi}{3} (2+3\cos \theta -\cos^3 \theta) \left( \gamma R^2   \left[ 1 + 2\ln\left(\frac{P+\frac{2\gamma}{R}}{k_B T \zeta n_1^{eq}}\right)+2\frac{\mu_l^0}{k_BT} \right] - R^3P\left[ 1- \ln\left(\frac{P+\frac{2\gamma}{R}}{k_B T \zeta n_1^{eq}}\right)-\frac{\mu_l^0}{k_BT} \right] \right)\,.
    \label{eq:energy_diff4}
\end{align}
Using \eqnref{n_eq} we can rewrite the last expression as 
\begin{equation}
    \Delta G \approx \frac{\pi}{3} (2+3\cos \theta -\cos^3 \theta) \left( \gamma R^2   \left[ 1 + 2\ln\left(\frac{P+\frac{2\gamma}{R}}{P_{a} \zeta}\right)\right] - R^3P\left[ 1- \ln\left(\frac{P+\frac{2\gamma}{R}}{P_{a} \zeta}\right) \right] \right)\,.
    \label{eq:energy_diff4-2}
\end{equation}
We note that $\theta=0^{\circ}$ corresponds to the case of bubble formation in the bulk, and \eqnref{energy_diff4-2} reduces to
\begin{equation}
    \Delta G \approx \frac{4}{3}\pi  \left( \gamma R^2   \left[ 1 + 2\ln\left(\frac{P+\frac{2\gamma}{R}}{P_{a} \zeta}\right)\right] - R^3P\left[ 1- \ln\left(\frac{P+\frac{2\gamma}{R}}{P_{a} \zeta}\right) \right] \right)\,.
    \label{eq:energy_diff_theta0}
\end{equation}

\section{Results and discussions}
\begin{figure}[t!]
\centering
\begin{subfigure}{.45\textwidth}
		\includegraphics[width= 0.95\textwidth]{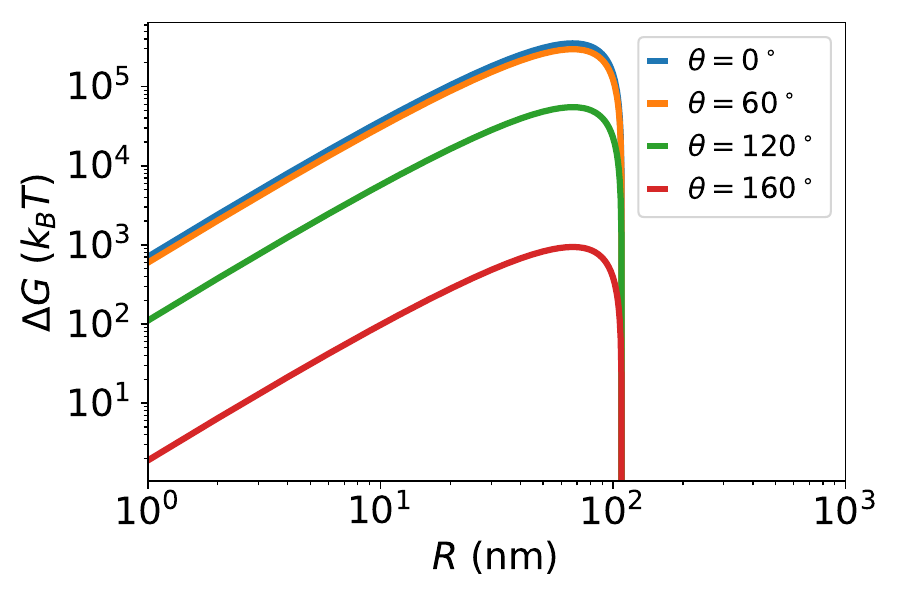}
		\subcaption{}
		\label{fig:varytheta_R}
	\end{subfigure}
   \begin{subfigure}{.45\textwidth}
		\includegraphics[width= 0.95\textwidth]{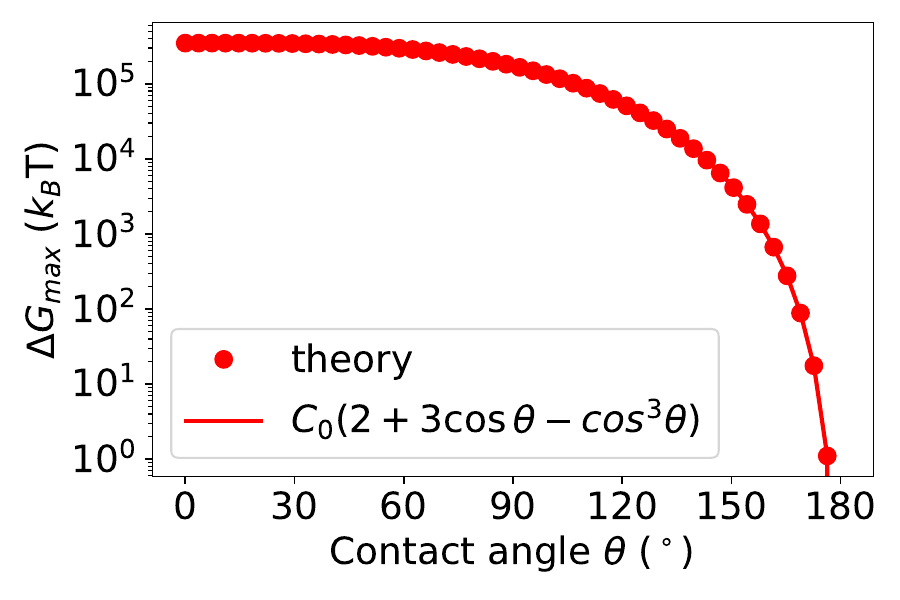}
		\subcaption{}
		\label{fig:max_e}
	\end{subfigure}
\caption{(a) Energy difference $\Delta G$ as a function of bubble radius $R$ at supersaturation $\zeta=100$ for different contact angles 
$\theta=0^{\circ}$ (blue), $\theta=60^{\circ}$ (orange), $\theta=120^{\circ}$ (green), $\theta=160^{\circ}$ (red). (b) Activation energy as a function of the contact angle $\theta$ (symbols). The solid line represents the function $\Delta G_{max} = C_0(2+3\cos \theta-\cos^3\theta)$, where $C_0=\Delta G_{max}^{\theta=0^{\circ}}$ is the activation energy at contact angle $\theta=0^{\circ}$.}
\label{fig:energy_diff}
\end{figure}

In the following, we apply our free-energy analysis to quantitatively predict the energy barrier and critical nucleation radius of a bubble at a catalyst surface. 
We take the oxygen evolution reaction in water as an example, a common phenomenon in PEMWE. The oxygen produced during PEMWE operation is normally released into the air, where the partial pressure of oxygen in air is about $P_a=2.1\times10^4\ Pa$.
In addition, we assume $k_B=1.38 \times 10^{-23}\ J/K$, the surface tension of water $\gamma_{wg}=0.072\ N/m$, the pressure  $P=1.01325\times10^5\ Pa$ and room temperature at $T=293\ K$.
\figref{varytheta_R} shows the dependence of the energy difference $\Delta G$ (\eqnref{energy_diff4-2}) on the bubble radius $R$
at supersaturation $\zeta=100$ for different contact angles $\theta=0^{\circ}$ (blue), $\theta=60^{\circ}$ (orange), $\theta=120^{\circ}$ (green), $\theta=160^{\circ}$ (red). For all the contact angles, the energy difference increases at small bubble radii, reaches a maximum at $R\approx 80\ nm$, and then decreases sharply to zero at a critical nucleus size $R_c\approx 109\ nm$. We note that \figref{varytheta_R} is presented in a log-log scale, and the further decrease of the energy difference beyond $R_c$ is therefore not shown. 
As shown in Fig.~\ref{fig:energy_diff}, while the critical nucleation radius of the bubble is independent of the contact angle, the activation energy (maximum energy difference) has a strong dependence on the contact angle, which is in qualitative agreement with previous work~\cite{Vachaparambil_2018}. 
By taking the maximum of \eqnref{energy_diff4-2}, we obtain the activation energy as a function of contact angle $\theta$, as shown in \figref{max_e}. At $\theta=0^{\circ}$, the activation energy takes its maximum, $\Delta G_{max}\approx 1.41\times 10^{-15}\ J \approx 3.49 \times 10^5\ k_BT$. This demonstrates that bubble nucleation in the bulk ($\theta=0^{\circ}$) requires a higher activation energy than heterogeneous nucleation, in agreement with previous work~\cite{VolmerWeber1926,Farkas1927,KELTON201085,Lubetkin1995,Vachaparambil_2018}. 
The activation energy decreases with increasing contact angle, following the relation $\Delta G_{max}^{\theta} = \Delta G_{max}^{\theta=0^{\circ}}(2+3\cos \theta-\cos^3\theta)$, as encoded in the functional form of \eqnref{energy_diff4-2}. At $\theta=180^{\circ}$ the activation energy vanishes, indicating the formation of a gas film at the solid surface. 

\begin{figure}[b!]
\centering
 \begin{subfigure}{.31\textwidth}
		\includegraphics[width= 0.99\textwidth]{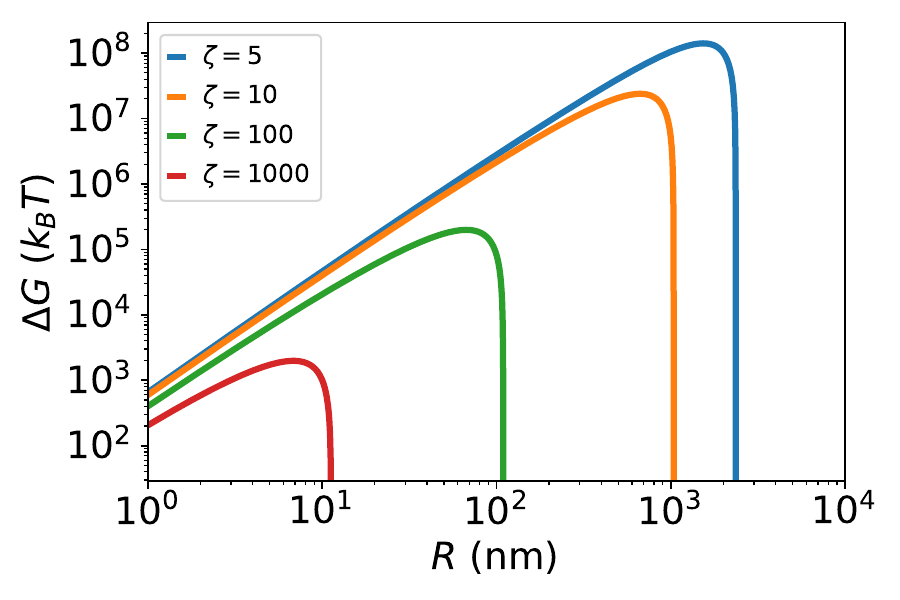}
		\subcaption{}
		\label{fig:varyzeta-R}
	\end{subfigure}
 \begin{subfigure}{.31\textwidth}
		\includegraphics[width= 0.99\textwidth]{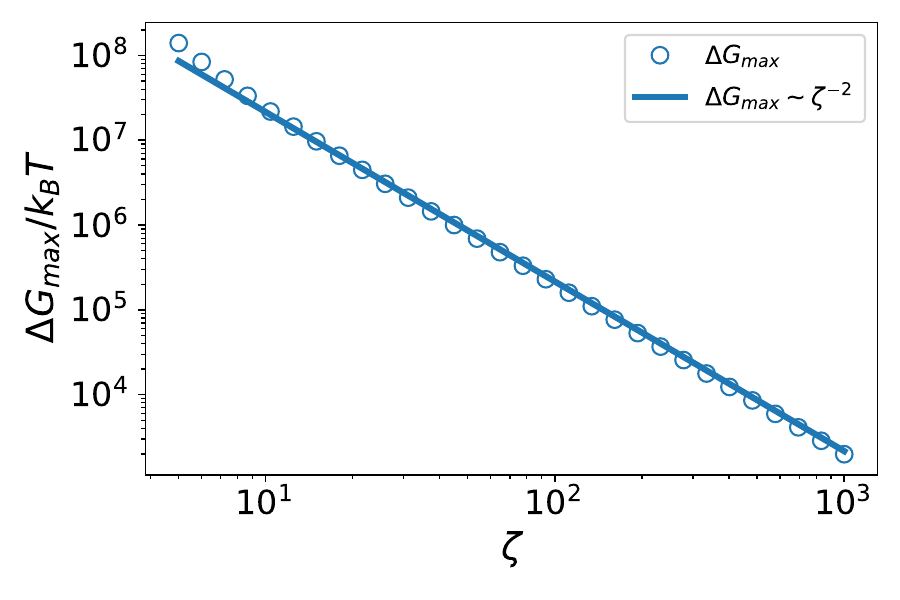}
		\subcaption{}
		\label{fig:varyzeta-Rc}
	\end{subfigure}
  \begin{subfigure}{.31\textwidth}
		\includegraphics[width= 0.99\textwidth]{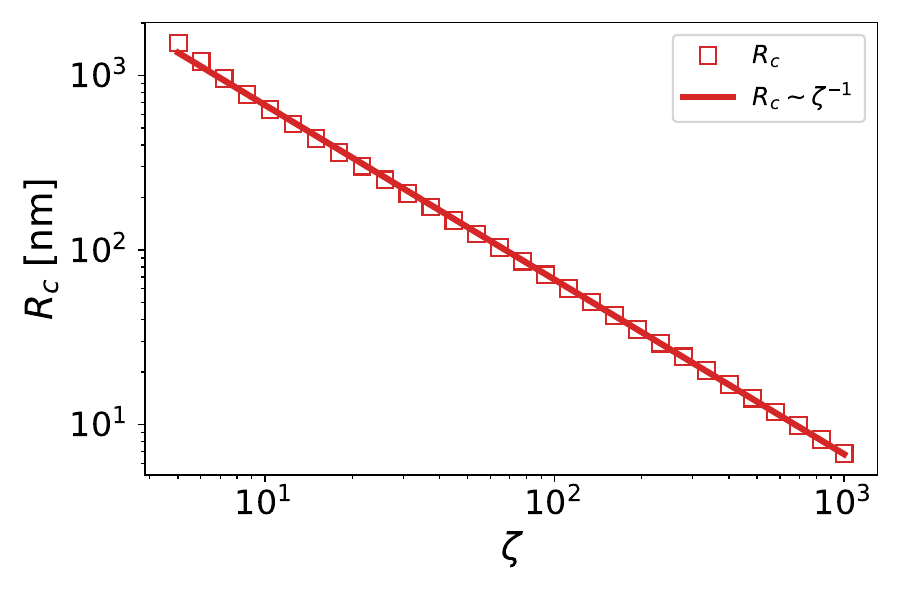}
		\subcaption{}
		\label{fig:varyzeta-Rc}
	\end{subfigure}
\caption{(a) Energy difference $\Delta G$ as a function of bubble radius 
at contact angle $\theta=85^{\circ}$ for different supersaturations $\zeta=5$ (blue), $\zeta=10$ (orange), $\zeta=100$ (green), $\zeta=1000$ (red). 
(b) Activation energy $\Delta G_{max}$ and (c) critical nucleus size $R_c$ as a function of the supersaturation $\zeta$.}
\label{fig:energy_diff-zeta}
\end{figure}

Based on \eqnref{energy_diff4-2}, the activation energy and the critical nucleus size are expected to depend strongly on the supersaturation $\zeta$. For a fixed contact angle of $\theta = 85^{\circ}$, \figref{varyzeta-R} shows the energy difference $\Delta G$ as a function of bubble radius $R$ for different supersaturation levels. Similar to \figref{varytheta_R}, the energy difference initially increases at small bubble radii, reaches a maximum, and then rapidly drops to zero at the critical nucleus size.
However, both the activation energy and the critical nucleus size decrease with increasing supersaturation $\zeta$, in line with the previous work~\cite{VolmerWeber1926,Farkas1927,KELTON201085,Lubetkin1995,Vachaparambil_2018,gadea2023}.
In~\figref{varyzeta-Rc} we show the dependence of the activation energy (circles) and the critical nucleus size (squares) on supersaturation. Surprisingly, the activation energy decreases with increasing $\zeta$ according to a power law, $\Delta G_{max} \sim \zeta^{-2}$, while the critical nucleus size follows $R_c \sim \zeta^{-1}$. 

Our theoretical analysis is generally applicable to other solvent-gas systems, such as the nucleation of $H_2$ and $N_2$ bubbles. 
Edwards et al. experimentally characterized the nuclei of single nanobubbles in a system 
where electrochemical generation creates a gas supersaturation at a nanoelectrode~\cite{Edwards2019}.
For hydrogen bubbles at atmospheric pressure ($P = 1\ \mathrm{atm}$) and room temperature ($T = 293\ \mathrm{K}$), the reported nucleation radius is $r_c \approx 4.7\ \mathrm{nm}$ for a contact angle $\theta = 146^\circ$ and supersaturation $\zeta = 312$, with an activation energy $\Delta G_{\text{max}} \approx 34\ k_B T$.
In the case of nitrogen bubbles under the same $P$ and $T$, the measured nucleation radius is $r_c \approx 8.1\ \mathrm{nm}$ for $\theta = 154^\circ$ and $\zeta = 178$, yielding $\Delta G_{\text{max}} \approx 20\ k_B T$. Using our theoretical model (\eqnref{energy_diff4-2}) with $\zeta = 312$, $P_a = P = 1.01325 \times 10^5\ \mathrm{Pa}$, and $\theta = 146^\circ$, we obtain a critical radius $r_c \approx 4.5\ \mathrm{nm}$ and activation energy $\Delta G_{\text{max}} \approx 35.6\ k_B T$ for hydrogen—in close quantitative agreement with experiment.
Similarly, for nitrogen with $\zeta = 178$ and $\theta = 154^\circ$, our model predicts $r_c \approx 8.1\ \mathrm{nm}$ and $\Delta G_{\text{max}} \approx 19.1\ k_B T$, also consistent with the experimental results.  Furthermore, for oxygen with $\zeta \approx 168$ and $\theta = 154^\circ$, our model predicts $r_c \approx 8.5\ \mathrm{nm}$ and $\Delta G_{\text{max}} \approx 41\ k_B T$, in good agreement with the experimental report~\cite{Edwards2019, Soto2018}. 

\begin{figure}[t!]
    \centering
        \begin{subfigure}{.333\textwidth}
		\includegraphics[width= 0.95\textwidth]{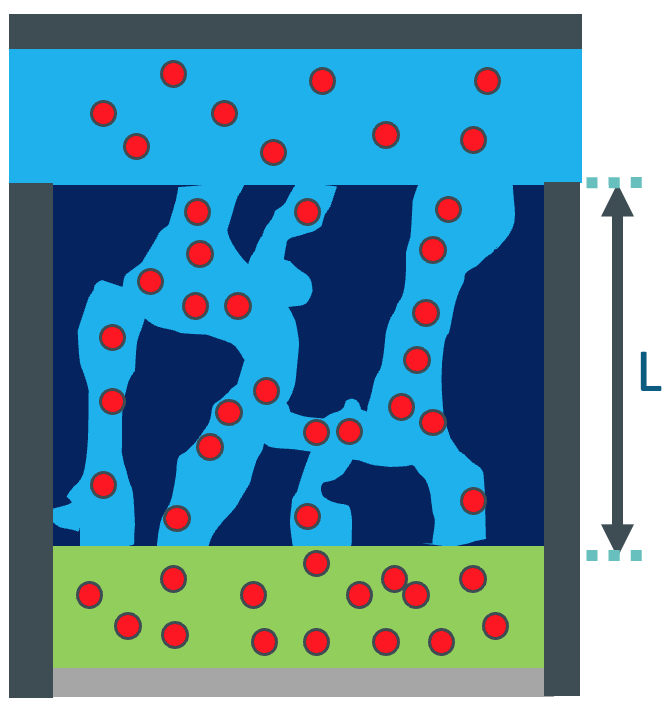}
		\subcaption{}
		\label{fig:snap-pewme}
	\end{subfigure}
 \hspace*{5mm}
	\begin{subfigure}{.3\textwidth}
		\includegraphics[width= 0.9\textwidth]{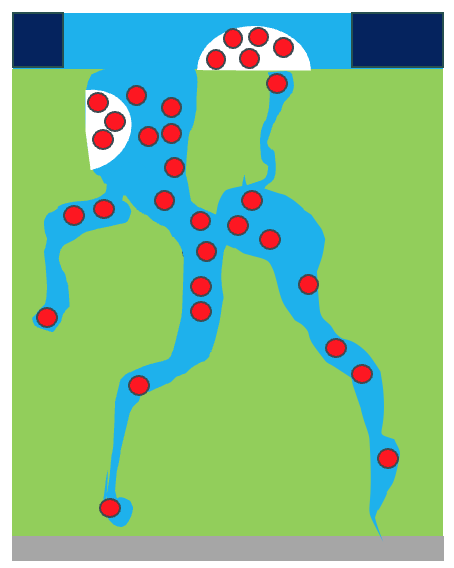}
		\subcaption{} 
		\label{fig:snap-pores-CL}
	\end{subfigure}
    \caption{a) Schematic illustration of the anode side of PEMWE. The gas (red circles) diffuses away through the catalyst layer (light green) and the porous transport layer (dark blue) to a water
channel (light blue). The bottom part (light grey) represents the membrane. The side and the top parts (dark grey) denote the bipolar plates. b) Sketch of bubble formation and gas transport in a catalyst layer. Bubbles (white) preferentially nucleate in the larger pores and at the CL–PTL interface, while in smaller pores, gas is generated and diffuses toward larger pores and the CL–PTL interface.
}
    \label{fig:snap-bubble}
\end{figure}

For \revisedtext{the} given values of temperature, pressure, and supersaturation, our theoretical model can be employed to predict the activation energy and the critical nucleus size. 
However, in practice, it is highly challenging to directly measure the degree of supersaturation during electrolyzer operation. 
We therefore estimate the maximum supersaturation that can be achieved at a specified current density\revisedtext{. This provides} a practical link between \revisedtext{the} experimentally accessible operating parameters 
and the nucleation characteristics predicted by our model.
We assume that before bubble nucleation, the generated gas at the catalyst layer is removed by diffusion only.
The gas flux in units of $kg/(m^2 \cdot s)$ generated by the electrochemical reaction is written as
\begin{equation}
   J_{e} = \frac{i M_o}{n_e F} ,
\end{equation}
where $i$ is the electric current density, $n_e=4$ is the number of electrons needed for the production of an oxygen molecule from water, $F$ is the Faraday constant, and $M_o$ is the molar mass of the gas. 
In PEMWE, when bubbles are absent, the gas diffuses away through the catalyst layer and the porous transport layer to a water channel~\cite{sangtam2023a,SELAMET20135823}, as illustrated in~\figref{snap-pewme}.
Following Fick's law, the mass flux of oxygen diffusion through the water solution can be estimated as
\begin{equation}
    J_d = -D \Delta c = -D \frac{c_0-c_\infty}{L},
\end{equation}
where $D$ is the diffusion coefficient of gas in water, $c_\infty$ and $c_0$ are the concentration\revisedtext{s} of gas in water in the channel and at the catalyst layer, and $L$ is the characteristic length. 
The total mass flux is the product of flux and surface area of the catalyst layer.
We note that the current density $i$ is defined as the total current through the electrolyzer divided by the geometrical surface area of the catalyst layer, effectively averaging over the CL depth (i.e., the thickness of the CL). 
Furthermore, we assume that the surface area of the catalyst layer represents the effective area for diffusion between the CL and the water channel, as shown in \figref{snap-pewme}.
Accordingly, we directly compare the reaction flux with the diffusion flux and determine
the maximal achievable supersaturation, $\zeta_m=c_0/c_s$, at a given current density as
\begin{equation}
    \zeta_m = \left( \frac{i LM_o}{Dn_e F}+c_\infty \right) / c_s,
    \label{eq:zeta_m}
\end{equation}
where $c_s$ is the saturation concentration of gas in water. 
Again, we consider the oxygen evolution reaction in PEWME as an example case. We take the diffusivity of the oxygen molecule in water as $D = 2\times10^{-9}\ m^2/s$~\cite{JAMNONGWONG2010758},
the Faraday constant as $F = 96485\ C/mol $, and the molar mass of oxygen as $M=32\ g/mol$. The water in the flow channel is assumed to contain a saturated oxygen concentration $c_\infty=9.1 \times 10^{-3}\ kg/m^3$ at $1$ atm at $T=293\ K$. 
The oxygen needs to diffuse through the PTL to reach the water channel, and we take 
the characteristic diffusion length at the order of $D_L = 2.5\times 10^{-5}\ m$, corresponding to the thickness of a thin PTL~\cite{MO2016817}.
Typically, PEMWE operates at a current density of order $1\ A/cm^2$~\cite{shiva_kumar_hydrogen_2019}, 
which may trigger a supersaturation level of $\zeta_m \approx 1000$ based on \eqnref{zeta_m}, resulting in a critical nucleus radius of $R_c\approx 8\ nm$ (from \eqnref{energy_diff4-2}). The predicted supersaturation level is in agreement with experimental measurements that the supersaturation may lie in the range $100-1000$ before bubble nucleation~\cite{SAKUMA20147638,MATSUSHIMA20095858,Luo2019-bubble}. We note that $\zeta_m \approx 1000$ is the maximal theoretically achievable supersaturation under the specified current density $1A/cm^2$ and associated conditions (e.g., a thin PTL). In practice, however, nucleation can occur well before this supersaturation is reached, and the value of $\zeta_m$
 may vary across experiments employing different current densities or PTLs of varying thickness.

The porous structure of the catalyst layer typically exhibits a characteristic pore size distribution, ranging from a few nanometers up to about $1000\ nm$~\cite{LV2023232878}.
Based on our free energy analysis, we propose a possible description for gas generation, bubble formation, and transport within the catalyst layer. Electrochemical reactions occur at the catalyst surfaces of all pores in contact with water in PEMWE.
The local reaction rate may vary depending on the ionic conductivity of the ionomer phase~\cite{bernt2016,MA2025107965} and the electronic conductivity of the catalyst layer~\cite{bernt2021,MA2025107965}. Bubbles preferentially nucleate in the larger pores and at the CL–PTL interface, while at the smaller pores, gas is generated and diffuses toward bigger pores and the CL–PTL interface, as depicted in~\figref{snap-pores-CL}. 
These proposed descriptions are consistent with the experimental observations reported by Yuan et al.\cite{YUAN2024}: once electrolysis ceases, residual gas in small pores and bubbles in large pores (if any) in the CL continue to migrate toward the CL–PTL interface, where larger bubbles eventually form. Similarly, the findings of Mo et al. \cite{Mo2016} can be explained as follows: the reaction rate is much higher near the PTL than farther away. In regions near the PTL, the reaction flux exceeds the diffusion flux, leading to a rapid increase in local gas supersaturation. This, in turn, it triggers bubble nucleation at the CL–PTL interface near the PTL. Another possible explanation for bubble formation next to the PTL is heterogeneous nucleation, which may occur if the PTL surface is less water-wettable.

\revisedtext{Our theoretical model, \eqnref{energy_diff4-2}, describes the energy difference before and after bubble nucleation as a function of surface tension, nanobubble radius, pressure, partial pressure, and supersaturation. To the best of our knowledge, no experimental studies have systematically investigated the energy difference while independently varying all of these parameters. Consequently, a direct comparison between \eqnref{energy_diff4-2} and experimental measurements is not feasible. Nevertheless, we assess the validity of our model by comparing its predictions with available experimental data on the nucleation of hydrogen, nitrogen, and oxygen nanobubbles at given contact angles and supersaturations. This comparison shows quantitative agreement, supporting the applicability of our approach. Furthermore, our predictions are consistent with molecular dynamics simulations~\cite{Perez2019} of surface nanobubble nucleation, and show qualitative agreement with classical density functional theory results~\cite{Oxtoby1988}.
}

\revisedtext{We note several limitations of our model that may affect the accuracy and generality of the results. As discussed in Sec.~\ref{sec:analysis}, the ideal gas approximation assumes perfect gas behaviour and may break down at the (sub-)nanoscale, where molecular interactions and sizes become significant. In addition, the assumption of constant surface tension at very small radii may not adequately capture the interplay between surface energy and curvature. Furthermore, our simplified treatment of supersaturation does not account for its spatial and temporal variations.
Future improvements could include the use of more realistic equations of state (e.g., the van der Waals equation of state) to account for intermolecular interactions and molecular sizes, the incorporation of curvature-dependent surface tension, and the development of a more comprehensive description of supersaturation that captures its spatial and temporal evolution.
}

\section{Conclusions}
In summary, we present a free-energy analysis of bubble nucleation on a flat catalyst surface. 
The analysis provides quantitative predictions for both the activation energy barrier and the critical nucleus size as functions of 
supersaturation, temperature, and pressure, while also incorporating the influence of surface wettability.
Our analysis reveals that the activation energy, $\Delta G_{max}$, decreases with increasing supersaturation $\zeta$, following a power-law scaling of $\Delta G_{max}\sim \zeta^{-2}$, which has not been reported to the knowledge of the authors. Meanwhile, the critical nucleus radius is found to scale inversely with supersaturation as $R_c\sim \zeta^{-1}$. 
Our theoretical predictions for the critical nucleation radius show quantitative agreement with the experimental results for hydrogen, oxygen and nitrogen nanobubbles~\cite{Edwards2019}. Beyond the thermodynamic description, we further propose a simplified model that couples gas diffusion with electrochemical reaction kinetics to estimate the maximum achievable gas supersaturation as a function of the current density. This enables the prediction of nucleation characteristics under the operating conditions of electrochemical devices. 

In this work, we focused solely on bubble nucleation on flat surfaces. However, our methodology can be extended directly to investigate the bubble nucleation at curved surfaces~\cite{Baidakov2013} or in cavities~\cite{JONES199927,Atchley1989}. Furthermore, our work can incorporate variations in surface tension at the nanoscale, which may become significant for radii below approximately $10\ nm$~\cite{bartell2001,german2016a}.

\begin{acknowledgements}
We acknowledge financial support from
the German Federal Ministry of Education and Research (BMBF) -- Project H2Giga/AEM-Direkt (Grant number 03HY103HF),
and the Bavarian Ministry for Economy, State Development and Energy -- Project H2Season (Grant number RMF-SG20-3410-6-10-11).
\end{acknowledgements}

\section*{Data availability}
The data that support the findings of this study are openly available at
\href{http://doi.org/10.5281/zenodo.19084926}{10.5281/zenodo.19084926}. 


\section*{Conflicts of interest}
The authors declare no competing financial interest.

\bibliographystyle{unsrt}
\bibliography{Ref-Xie.bib}
\end{document}